\documentclass[journal]{IEEEtran}
\ifCLASSINFOpdf \else \fi

\usepackage{booktabs} 
\usepackage{multirow}
\usepackage{placeins}
\usepackage{tabularx}  
\usepackage{array}      
\usepackage{booktabs}   
\usepackage{longtable}  
\usepackage{booktabs}
\usepackage{lscape}
\usepackage{bbold}
\usepackage{ifpdf}
\usepackage{xcolor}
\usepackage{ulem} 
\usepackage{subcaption}%
\usepackage[cmex10]{amsmath}
\usepackage{amssymb}
\usepackage{verbatim}
\usepackage{algorithm}
\usepackage{algpseudocode}
\usepackage{amsmath}
\usepackage{amsfonts}
\usepackage{array}
\usepackage{latexsym}
\usepackage{color}
\usepackage{textgreek}
\usepackage{subscript}
\usepackage{rotating}
\usepackage{booktabs,multirow}
\usepackage{mdframed}	
\usepackage{tabularx}
\usepackage{amsmath}
\usepackage{makecell}
\usepackage{tabto}
\usepackage{mathrsfs}
\usepackage{url}
\usepackage{cite}
\usepackage{listings}
\usepackage{array}
\usepackage[table]{xcolor}
\usepackage{ragged2e}
\usepackage{booktabs}

\definecolor{bluepurple}{rgb}{0.4, 0.1, 0.9} 
\newcommand{\notemahsanew}[1]{\textcolor{black}{#1}}

\definecolor{pinkpurple}{rgb}{0.6, 0.1, 0.9} 

\usepackage{graphicx}

\usepackage[cmex10]{amsmath}
\usepackage{amssymb}
\usepackage{bbold}
\usepackage{float}
\usepackage[cmex10]{amsmath}
\usepackage{amssymb}
\usepackage{verbatim}
\usepackage{array}
\usepackage{latexsym}
\usepackage{color}
\usepackage{tabularx} 
\usepackage{textgreek}
\usepackage{subscript}
\usepackage{rotating}
\usepackage{booktabs,multirow}
\usepackage{mdframed}	
\usepackage{tabularx}
\usepackage{amsmath}
\usepackage{makecell}
\usepackage{tabto}

\usepackage{mathrsfs}
\usepackage{url}
\usepackage{cite}
\usepackage{listings}
\usepackage{array}
\usepackage[table]{xcolor}

\definecolor{pinkpurple}{rgb}{0.1, 0.3, 0.9} 
\usepackage{array}
\usepackage{algorithm}
\usepackage{soul}
\sethlcolor{green}
\usepackage{url}
\usepackage{pifont}
\usepackage{xcolor}
\newcommand{\cmark}{\textcolor{green}{\ding{51}}} 
\newcommand{\xmark}{\textcolor{red}{\ding{55}}}   
\usepackage{multirow}
\usepackage{graphicx} 
\begin{document}
\title{On-Dyn-CDA: A Real-Time Cost-Driven Task Offloading Algorithm for Vehicular Networks with Reduced Latency and Task Loss}

\author{
	Mahsa~Paknejad,~Parisa~Fard~Moshiri,~Murat~Simsek~\IEEEmembership{Senior~Member,~IEEE},\\~Burak Kantarci,~\IEEEmembership{Senior Member,~IEEE}~and~Hussein~T.~Mouftah,~\IEEEmembership{Fellow,~IEEE}
	\thanks{
	The authors are with the School of Electrical Engineering and 
        Computer Science at the University of Ottawa, Ottawa, ON, K1N 6N5, Canada.
        E-mail: \{mahsa.paknejad, parisa.fard.moshiri, murat.simsek, burak.kantarci, mouftah\}@uottawa.ca}   
}

\markboth{IEEE Transactions on Internet of Things}{}

\maketitle
 \thispagestyle{empty}
  \pagestyle{empty}
\begin{abstract}
Real-time task processing is a critical challenge in vehicular networks, where achieving low latency and minimizing dropped task ratio depend on efficient task execution. Our primary objective is to maximize the number of completed tasks while minimizing overall latency, with a particular focus on reducing number of dropped tasks.
To this end, we investigate both static and dynamic versions of an optimization algorithm. The static version assumes full task availability, while the dynamic version manages tasks as they arrive. We also distinguish between online and offline cases: the online version incorporates execution time into the offloading decision process, whereas the offline version excludes it, serving as a theoretical benchmark for optimal performance.
We evaluate our proposed Online Dynamic Cost-Driven Algorithm (On-Dyn-CDA) against these baselines. Notably, the static Particle Swarm Optimization (PSO) baseline assumes all tasks are transferred to the RSU and processed by the MEC, and its offline version disregards execution time, making it infeasible for real-time applications despite its optimal performance in theory.
Our novel On-Dyn-CDA completes execution in just 0.05 seconds under the most complex scenario, compared to 1330.05 seconds required by Dynamic PSO. It also outperforms Dynamic PSO by 3.42\% in task loss and achieves a 29.22\% reduction in average latency in complex scenarios. Furthermore, it requires neither a dataset nor a training phase, and its low computational complexity ensures efficiency and scalability in dynamic environments.
\end{abstract}

\IEEEpeerreviewmaketitle

\begin{table*}[!t]
\caption{A comparison of our work with existing literature}
\centering
\scriptsize
 \scalebox{0.8}{ 
\renewcommand{\arraystretch}{1.4} 
\newcolumntype{M}[1]{>{\centering\arraybackslash}m{#1}}
\begin{tabular}{|M{0.8cm}|M{0.8cm}|M{0.8cm}|M{1cm}|M{0.8cm}|M{0.8cm}|M{1.5cm}|M{1.5cm}|c|c|M{1.6cm}|}
\hline

\textbf{Paper} & \textbf{\# Tasks} & \textbf{\# MEC Servers} & \textbf{Dynamic} & \multicolumn{2}{c|}{\textbf{Training}} & \textbf{Simulation Tool} & \multicolumn{3}{c|}{\textbf{Key Performance Indicator (KPI)}} & \textbf{Algorithm} \\ \cline{5-6} \cline{8-10}
               &                   &                         &                   & \textbf{Offline} & \textbf{Online} &                          &                           \textbf{Algorithm Execution Time} & \textbf{Latency} & \textbf{Dropped Task Ratio} & \\
\hline

\cite{iot6}  & 40-120 & 10 & {\xmark} & - & - & - & {\xmark} &  \multicolumn{1}{c|}{\cmark} & \multicolumn{1}{c|}{\xmark} & ACO \\ \hline

\cite{iot1}  & - & 10 & {\xmark} & - & - & Python & {\xmark} & \multicolumn{1}{c|}{\cmark} & \multicolumn{1}{c|}{\xmark}  & PBTSA, PBLA, TBTOA \\ \hline

\cite{iot2}  & - & 6 & {\xmark} & - & - & - & {\xmark} & \multicolumn{1}{c|}{\cmark} & \multicolumn{1}{c|}{\cmark}  & MCLA \\ \hline

\cite{iot3}  & - & 1 & {\xmark} & - & - & MATLAB & {\xmark} &  \multicolumn{1}{c|}{\cmark} & \multicolumn{1}{c|}{\cmark} & Theoretical analysis \\ \hline

\cite{iot5}  & 50 & 4 & {\xmark} & - & - & MATLAB & {\xmark} &  \multicolumn{1}{c|}{\cmark} & \multicolumn{1}{c|}{\cmark} & Theoretical analysis \\ \hline

\cite{iot4}  & - & - & {\cmark} & {\xmark} & {\cmark} & SUMO + MATLAB & {\xmark} &  \multicolumn{1}{c|}{\cmark} & \multicolumn{1}{c|}{\xmark} & LSTM, PPO \\ \hline

\cite{iot7}  & - & 4 & {\cmark} & {\xmark} & {\cmark} & Python & {\xmark} &  \multicolumn{1}{c|}{\cmark} & \multicolumn{1}{c|}{\xmark} & MADDPG \\ \hline

\cite{iot8}  & - & - & {\cmark} & {\xmark} & {\cmark} & Python & {\xmark} &  \multicolumn{1}{c|}{\cmark} & \multicolumn{1}{c|}{\xmark} & DRL \\ \hline

\cite{iot9}  & - & - & {\cmark} & {\cmark} & {\xmark} & DAGGEN + Python & {\xmark} &  \multicolumn{1}{c|}{\cmark} & \multicolumn{1}{c|}{\xmark} & SMRL-MTO \\ \hline

\cite{iot10}  & - & 4, 32 & {\cmark} & {\cmark} & {\xmark} & real historical dataset & {\xmark} &  \multicolumn{1}{c|}{\cmark} & \multicolumn{1}{c|}{\xmark} & RL \\ \hline

Ours  & 50-100-200 & 2 & {\cmark} & - & - & SUMO + python & {\cmark} &  \multicolumn{1}{c|}{\cmark} & \multicolumn{1}{c|}{\cmark} & On-Dyn-CDA, Off-Sta-PSO, On-Sta-PSO, On-Dyn-PSO \\ \hline

\end{tabular}
} 
\label{tab:gap}
\end{table*}

\section{Introduction}
Mobile Edge Computing (MEC) has emerged as an essential component in the era of the Internet of Things (IoT), where the number of connected devices, such as smartphones, sensors, and smart home appliances, is rapidly increasing \cite{IoT-ieee6}, \cite{intro1}. This proximity reduces dependency on distant centralized cloud servers, significantly lowering latency \cite{intro2}, \cite{parisaicc}.

Building on the benefits of MEC, Vehicular Edge Computing (VEC) extends this paradigm to meet the specific demands of Connected and Autonomous Vehicles (CAVs) and intelligent transportation systems \cite{intro3,AVCIL2024100773}. As vital elements of the IoT ecosystem, CAVs generate vast amounts of data from sensors, which must be processed in real time \cite{IoT-ieee8}, \cite{intro4}.
Machine learning (ML) models play a crucial role in processing this sensor data, enabling CAVs to accurately make decisions that ensure safe and efficient driving \cite{IoT-ieee2}, \cite{intro5}. However, due to the immense computational demands, onboard processing alone is often insufficient to meet real-time requirements \cite{IoT-ieee9}, \cite{IoT-ieee5}, \cite{intro5.5}. VEC provides a solution by enabling vehicles to offload computationally intensive tasks to nearby edge servers \cite{IoT-ieee1}, \cite{intro9}.
CAVs also rely heavily on communication technologies, specifically Vehicle-to-Vehicle (V2V) and Vehicle-to-Everything (V2X) communication, to exchange critical information \cite{intro6}. V2V communication enables vehicles to share data, facilitating cooperative driving and collision avoidance \cite{IoT-ieee7}. Meanwhile, V2X extends this communication to include infrastructure, such as traffic lights and roadside units (RSUs), as well as other connected devices \cite{intro6.5}. Despite VEC's advantages, real-time vehicular environments face challenges as multiple vehicles offload tasks, causing server queues, delays, and dropped tasks. Optimizing task processing order is crucial to maximize successful task completion and minimize the latency \cite{IoT-ieee10}. Conventional scheduling methods may not be adequate for these dynamic and high-load environments, making optimization algorithms, such as Particle Swarm Optimization (PSO), a promising approach \cite{parisa2}. However, traditional optimization algorithms like PSO often require considerable time to converge, which is impractical for real-time applications \cite{IoT-ieee4}. Consequently, developing a fast and reliable scheduling algorithm is essential to ensure that task offloading remains efficient, even in highly dynamic and demanding vehicular networks \cite{intro8}. This need for rapid and real-time approaches underscores the importance of innovative techniques to maintain low latency and high reliability in VEC environments. 


The main contributions of this paper are as follows:
\begin{enumerate}
\item 
 Online Dynamic Cost-Driven Algorithm (On-Dyn-CDA) is proposed for efficient real-time task scheduling in vehicular networks. It adapts to continuous task arrivals and resource changes, making it suitable for practical use while handling many tasks with low computational cost.
\item 
We develop a dynamic optimization approach to schedule tasks within decision windows, including tasks arriving before MEC availability, while accounting for the algorithm's execution time in each decision window.
\item 
Our study defines the theoretical upper bound, Offline Static PSO (Off-Sta-PSO), which assumes that all tasks are available for optimization in advance, enabling ideal scheduling without accounting for the execution cost of the optimization itself, which would otherwise directly impact task processing. We also compare our model to a real-time worst-case scenario, Online Static PSO (On-Sta-PSO), without dynamic behavior, showing our proposed model performs close to the theoretical optimum and far better than the worst-case baseline.
\end{enumerate}

Based on our contributions, our findings indicate that the novel dynamic algorithm, On-Dyn-CDA, efficiently manages real-time task scheduling in vehicular networks by dynamically adapting to continuous task arrivals and fluctuating resource availability with reduced computational cost. Section II presents the literature review, while Section III outlines the methodologies under study. The performance evaluation is covered in Section IV, and Section V concludes the study.

\begin{figure*}[!hbt]
        \centering
        \includegraphics[width = 0.6\textwidth, trim=1.2cm 9cm 1.2cm 1.2cm,clip]{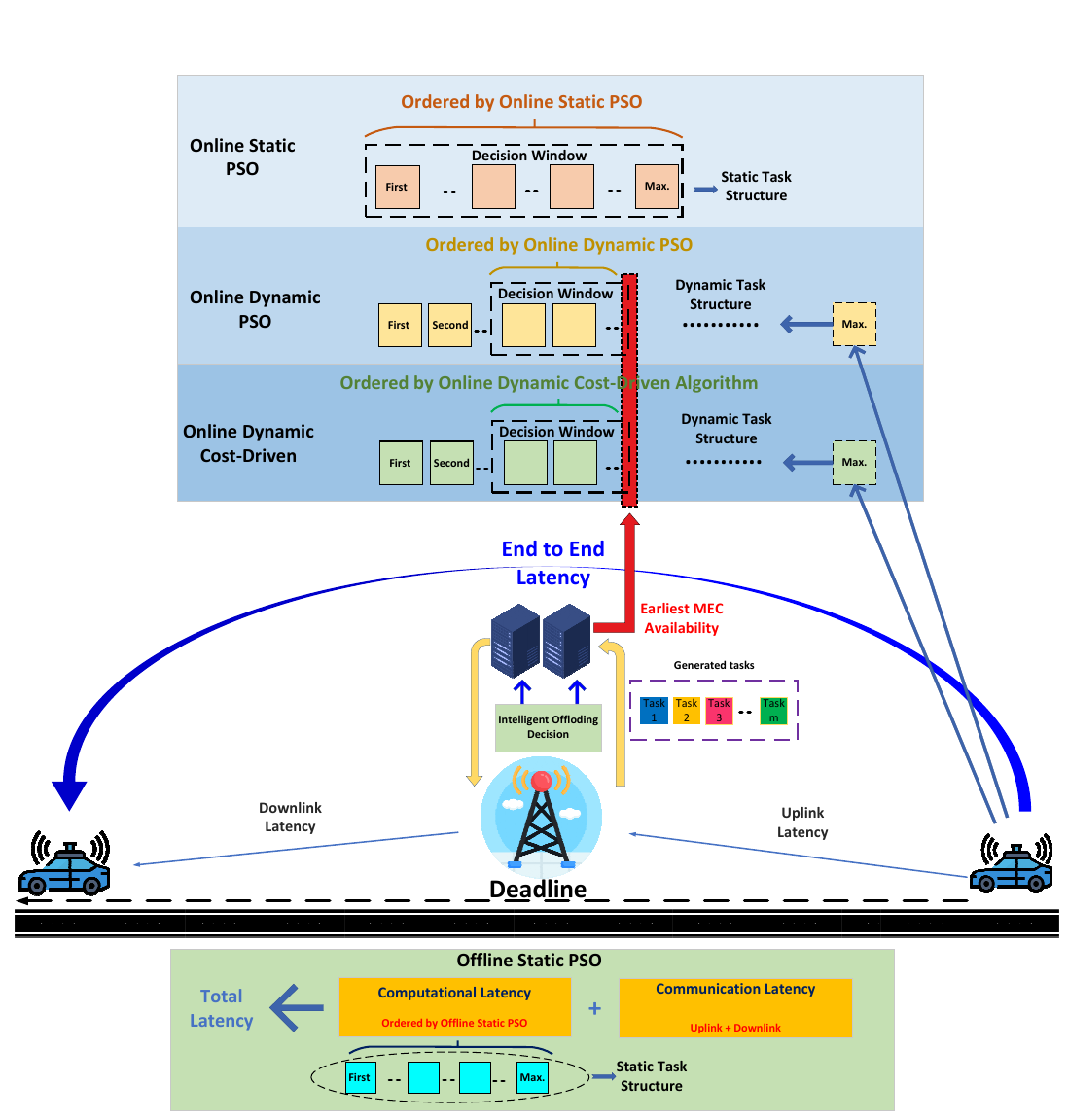}
        \caption{ General Concept of Vehicular Edge Computing for Static and Dynamic Task Offloading Process}
        \label{fig:scenario1} 
\end{figure*}

\begin{figure*}[!hbt]
        \centering
        \includegraphics[width = 0.76\textwidth, trim=1.2cm 2cm 1.2cm 1.2cm,clip]{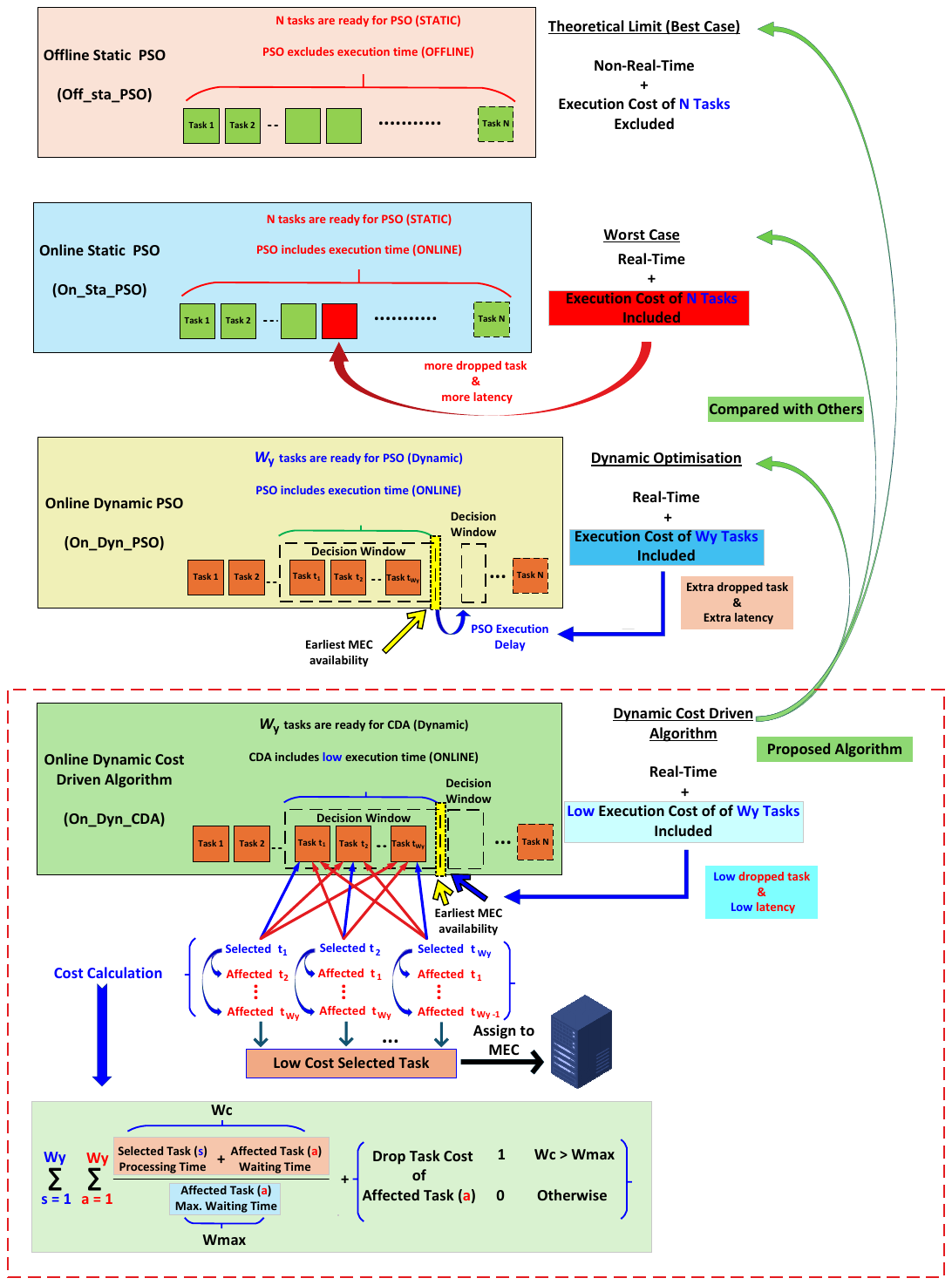}
        \caption{Comparative Analysis of the Proposed Algorithm and Existing Methods}
        \label{fig:scenario2} 
\end{figure*}
\section{Related Work}
Several studies have examined task offloading and resource allocation in MEC-enabled Internet of Vehicles (IoV) \cite{IoT-ieee3}, \cite{intro10}, highlighting the challenges of meeting low-latency requirements when relying solely on traditional cloud computing. Proposed solutions use multi-objective optimization, often targeting latency and load balancing. Ant colony optimization is one approach that explores offloading options while distributing load \cite{iot6}. Methods like multi-criteria decision-making and Simple Additive Weighing support comprehensive evaluation \cite{iot6}. 
Dependent tasks are critical in MEC-enabled 5G vehicular networks, as they comprise interlinked subtasks requiring sequential processing for system efficiency. In resource-limited settings, they can be modeled as Directed Acyclic Graphs (DAGs), necessitating effective scheduling to meet strict latency demands \cite{iot1}. This can be addressed using Priority-Based Task Scheduling Algorithm (PBTSA), which applies Reverse Breadth-First Search (RBFS) for subtask prioritization and a greedy offloading strategy to reduce overall delay \cite{iot1}. Dynamic communication switching improves task offloading in MEC-enabled vehicular networks by adapting to conditions using V2V and Vehicle-to-Infrastructure (V2I) modes. Switching between sub-6 GHz 5G NR and mmWave reduces delay and boosts execution, optimizing offloading based on mobility, contact time, and load. Reports show up to 1800 completed tasks, although the total is not stated \cite{iot2}.
\\
\indent Efficient resource allocation is vital for task offloading in MEC-enabled vehicular networks. Non-Orthogonal Multiple Access (NOMA) improves spectral efficiency through simultaneous transmissions, reducing congestion and boosting execution. Studies show task loss is more affected by vehicle arrival rates than speed, highlighting the importance of traffic density in allocation \cite{iot3}.
Building on NOMA's benefits, further optimizations allow multiple Vehicular User Equipments (VUEs) to share wireless resources, boosting spectrum use and system capacity. To handle the complexity of offloading and allocation, a decomposition strategy splits the problem into subproblems, solved with heuristics for efficient distribution, low latency, and balanced resource use \cite{iot5}.
Beyond spectrum and computation, task offloading must consider heterogeneous coverage with coexisting 5G base stations (BS)s and RSUs. In such scenarios, vehicles offload tasks based on proximity and real-time load, estimated using long short-term memory (LSTM) models. In one study, offloading is modeled as a constrained optimization problem, with Reinforcement Learning (RL) methods like Proximal Policy Optimization (PPO) improving decisions under dynamic conditions \cite{iot4}.
To further enhance offloading efficiency, distributed intelligence can dynamically balance workloads across edge resources. Multi-Agent Deep Deterministic Policy Gradient (MADDPG) algorithms optimize performance, reduce delays, and respect BS energy constraints \cite{iot7}. Task offloading effectiveness depends on task generation rate, user count, and simulation duration. However, \cite{iot7} omits the total number of tasks and simulation length, limiting result reproducibility. Building on RL, Actor-Critic-based Deep RL (DRL) optimizes task offloading by accounting for vehicle-RSU distance, channel conditions, and resource availability. It enables dynamic distribution, balancing latency, energy use, and resource allocation \cite{iot8}.
Seq2Seq-based Meta RL (SMRL-MTO) can also enhance offloading by capturing subtask dependencies and optimizing decisions. Using bidirectional Gated Recurrent Units (GRUs) with attention, it prioritizes key input sequences. In this research, the model handles DAGs with 12 subtasks across multiple servers, although exact numbers are unspecified \cite{iot9}. A model-agnostic meta-learning framework also enables fast adaptation to new MTO scenarios with minimal retraining \cite{iot9}.
To reduce the high exploration cost of online RL, offline RL is used for asynchronous task offloading in MEC, training a supervised model on historical data to simulate edge dynamics and optimize policies without real-world interaction \cite{iot10}. The paper demonstrates scalability by increasing the number of edge servers from 4 to 32, although the specific simulation tool used is not specified \cite{iot10}.

\indent Existing studies often overlook the combined impact of dynamic processing, task drop rates, and execution time, typically addressing these factors in isolation. However, low latency is less meaningful if many tasks are dropped. The priority should be meeting task deadlines, then minimizing latency for completed tasks. Additionally, fast algorithm execution is essential, as even optimal methods fail if they cannot make quick decisions. To address these challenges, we first analyze baselines, including the theoretical upper bound using PSO, and then implement a dynamic PSO that schedules tasks in real time, adapting to changing arrivals and resource availability. However, this method is inefficient for real-time use due to high execution time. To address this, we propose On-Dyn-CDA, which lowers task drop rates and latency while keeping execution time short. PSO is also evaluated in a static online setting to assess its computational cost across all tasks (worst-case scenario). These PSO models are implemented as baselines to compare with our proposed On-Dyn-CDA, a step rarely taken in existing studies.
Table I highlights the differences between our approach and the related studies.

\section{Methodologies under Study}



\subsection{Baseline Methods and the Proposed Algorithm}
For baselines, we employ \textbf{First-Come-First-Served (FCFS)} and \textbf{Shortest Deadline First (SDF)} \cite{parisa1}, both of which are greedy algorithms, along with \textbf{PSO}, a metaheuristic optimization technique that provides a more adaptive and efficient strategy for scheduling tasks in complex environments. In our approach, PSO determines the order in which tasks are
processed and assigns each task to a specific MEC server. This assignment is guided by the goal of reducing the number of dropped tasks and minimizing the overall latency for tasks that can be successfully processed. The algorithm evaluates key factors, such as task deadlines, server availability, and computational requirements, to ensure the efficient distribution of tasks across available resources.\\
\indent Based on \figurename \hspace{0.1pt}\ref{fig:scenario1} and \figurename \hspace{0.1pt}\ref{fig:scenario2}, three types of PSO algorithms are employed for comparison with our state-of-the-art approach: Off-Sta-PSO, On-Sta-PSO, and Online Dynamic PSO (On-Dyn-PSO). Each algorithm tackles task scheduling under different assumptions, ranging from ideal to realistic. Off-Sta-PSO represents the theoretical upper bound, with full prior knowledge of task arrivals and processing requirements before they reach the RSU. This reflects a static processing strategy, where the first two tasks start at their arrival times due to two available MECs. While real-world tasks arrive in real-time, making this assumption unrealistic, it still serves as a useful benchmark for evaluating our model.\\
\indent On-Sta-PSO offers a more realistic approach by computing waiting times in real-time, but still applies PSO to all tasks at once. It assumes all tasks must arrive at the RSU before scheduling begins, causing long waits for early tasks. As a result, in this worst-case scenario, tasks one and two start only after all tasks are received, delaying processing for PSO execution. In static algorithms, tasks one and two begin immediately using two available MECs, while later tasks are scheduled by PSO and processed sequentially.\\
\indent On-Dyn-PSO takes a dynamic, real-time approach, where PSO is executed multiple times as tasks are generated dynamically by Simulation of Urban MObility (SUMO). In this scenario, tasks one and two are immediately assigned to MEC servers upon arrival. Subsequent tasks arriving after task two are collected in a queue called the decision window. This queue includes the set of tasks that arrive between the start processing time of task two and the earliest MEC availability time. We should note that, in static scenarios, the decision window includes all tasks in the simulation, as we optimize the scheduling of all tasks together at once. The earliest MEC availability time is defined as the moment when at least one MEC becomes free to process tasks. Until just before this availability time, the PSO algorithm is applied to identify the optimal task for assignment to the first available MEC. After assigning the task, the availability time of each MEC is updated, and the earliest MEC availability time is determined by taking the minimum of the MECs’ availability times. A new decision window is created, ranging from the maximum start processing time of tasks across both MECs (corresponding to the start processing time of the most recently assigned task) to the new earliest MEC availability time. This decision window now includes all unassigned tasks that have already arrived or will arrive before the new earliest availability time. The PSO algorithm is then reapplied to identify the optimal task within this updated window, and the process continues iteratively until the simulation ends, ensuring that only one task is assigned to a MEC at a time. This approach enables dynamic, real-time task scheduling, closely reflecting how tasks would be managed in real-world scenarios. However, PSO requires significant execution time, making it impractical for real-world scenarios. To address this limitation, we propose a faster, real-time solution called the On-Dyn-CDA. This proposed algorithm is designed to handle tasks efficiently in real-time by leveraging a cost function to determine which task should be assigned to the available MEC first. Similar to On-Dyn-PSO, the first two tasks are automatically assigned to MEC1 and MEC2 upon arrival. Additionally, as in On-Dyn-PSO, tasks arriving after the start processing time of task two and the earliest availability time are gathered in a decision window. On-Dyn-CDA selects tasks based on the lowest computational cost, determined by a cost function detailed in the mathematical section.

\subsection{The mathematical framework for task offloading}
End-to-end latency, a key component of the objective function, depends on both computation and communication latencies. Each is explained separately to highlight its individual impact.

\subsubsection{computation latency}
The start processing time of each task is calculated by adding the start processing time of the preceding assigned task to its processing time. The start processing time is essential for determining a task’s waiting time, which contributes to its end-to-end latency. When task \( i \) is offloaded to a MEC server \( j \), it undergoes a delay composed of two main components: the time it waits before processing begins, and the actual processing duration. Based on (\ref{eq:waiting}), \(t_{i}^{\text{w}}\) represents the waiting time, indicating how long task \(i\) waits in the RSU before being processed. 

\begin{equation}
t_{i}^{\text{w}} = t_{i}^{\text{sp}} - t_{i}^{\text{ar}}
    \label{eq:waiting}
\end{equation} \( t_{i}^{sp} \) represents the start processing time of task \( i \), and \(t_{i}^{\text{ar}}\) denotes the arrival time at the RSU for task \(i\).
\tablename \hspace{0.1pt} \ref{tab:notations} provides an explanation of all the necessary notations for the mathematical formulas. The computation latency, \(L_{i}^{\text{p}}\), as expressed in (\ref{eq:Lcp}), is defined as the total of the processing time and the waiting time.

\begin{equation}
L_{i}^{\text{p}} = t_{i}^{\text{p}} + t_{i}^{\text{w}}
    \label{eq:Lcp}
\end{equation} 

\( t_{i}^{p} \) denotes the processing time of task \(i\) with actual processing times sourced from \cite{execution}.

\newcolumntype{M}[1]{>{\centering\arraybackslash}m{#1}}

\begin{table}[!t]
\centering
\renewcommand{\arraystretch}{1.3} 

\caption{Notation Table}
\label{tab:notations}
 \scalebox{0.8}{
{\footnotesize
\begin{tabular}{|>{\centering\arraybackslash}m{0.86cm}|m{5.4cm}|>{\centering\arraybackslash}m{2.4cm}|}
\hline
\textbf{Param.} & \textbf{Description} & \textbf{Eq.} \\
\hline
\( N\) & Total number of tasks & (\ref{eq:drop}) (\ref{eq:obj}) (\ref{eq:assign-con}) \\
\hline
\( M \) & Total number of MEC servers& (\ref{eq:drop}) (\ref{eq:obj}) (\ref{eq:ava}) (\ref{eq:assign-con})\\
\hline
\(N'\) &  Number of tasks ready for transmission to/from RSU simultaneously& (\ref{eq:B})\\
\hline
\({D} \) & Total number of dropped tasks& (\ref{eq:drop}) (\ref{eq:obj})\\
\hline
\( S_i \) & Size of task \( i \)& (\ref{eq:B}) (\ref{eq:Tcm})\\
\hline
\(x_{ij}\) & Binary variable indicating task \(i\) is assigned to MEC \(j\)& (\ref{eq:assign}) (\ref{eq:drop}) (\ref{eq:obj}) (\ref{eq:ava}) (\ref{eq:assign-con})\\
\hline
\(t_{i}^{\text{range}}\) & The time range from when the vehicle carrying task \(i\) enters the RSU’s coverage area until its task deadline & (\ref{eq:assign})\\
\hline
\(t_{i}^{\text{d}}\) & The deadline for processing task \( i \) and returning it to the vehicle& (\ref{eq:max_wait})\\
\hline
\( t_{i}^{\text{ar}} \) & Arrival time of task \( i \) to the RSU& (\ref{eq:waiting}) (\ref{eq:max_wait})\\
\hline
\(t_{i}^{\text{sp}}\) & Start processing time for task \( i \)& (\ref{eq:waiting}) (\ref{eq:ava})\\
\hline
\( L_{i}^{\text{e2e}} \) & End-to-end latency for task \( i \)& (\ref{eq:end2end}) (\ref{eq:assign}) (\ref{eq:obj})\\
\hline
\(L_{i}^{\text{p}} \) & Computation latency for task \( i \)& (\ref{eq:Lcp}) (\ref{eq:end2end})\\
\hline
\( T_{i}^{\text{p}}  \) & Processing time of task \( i \) & (\ref{eq:Lcp}) (\ref{eq:ava}) (\ref{eq:max_wait})\\
\hline
\(t_{i}^{\text{w}}\) & The time task \( i \)  waits in the RSU before starting processing & (\ref{eq:waiting}) (\ref{eq:Lcp})\\
\hline
\(t_{i}^{\text{w\_max}}\) & The maximum time that task \( i \) can wait before processing while meeting its deadline& (\ref{eq:max_wait}) (\ref{eq:d-cost}) (\ref{eq:final-cost-min})\\
\hline

\( L_i^{\text{tr}} \) & Transmission latency for task \(i\)& (\ref{eq:end2end})\\
\hline
\( t_i^{\text{tr}} \) & Transmission time for task \(i\)& (\ref{eq:Tcm})\\
\hline
\( r_i \) & Transmission rate for vehicle of task \( i \) & (\ref{eq:r}) (\ref{eq:Tcm})\\
\hline
\( B_i \) & Bandwidth of task \( i \) & (\ref{eq:B}) (\ref{eq:r})\\
\hline
\( B_{\text{max}}\) & Maximum bandwidth & (\ref{eq:B})\\
\hline
\( p \) & Transmission power & (\ref{eq:r})\\
\hline
\( g \) & Channel gain & (\ref{eq:r})\\
\hline
\( n_0 \) & Noise power& (\ref{eq:r})\\
\hline

\( t_{j}^{\text{av}} \) & The time when MEC \(j\) becomes available & (\ref{eq:ava}) (\ref{eq:ava-s})\\
\hline
\( t_{e}^{\text{av}} \) & Earliest MEC availability time& (\ref{eq:ava-s})\\
\hline
\(t_{s}^{\text{p}^{\text{}}}\) & The processing time of the selected task & (\ref{eq:d-cost}) (\ref{eq:final-cost-min})\\
\hline

\( d_{sa}^{\text{}} \) & Drop cost condition for an affected task based on the selected task& (\ref{eq:d-cost}) (\ref{eq:final-cost-min})
\\
\hline

\( S_{}^{\text{}} \) & The set of tasks available within the decision window& (\ref{eq:s}) (\ref{eq:q})
\\
\hline

\( Q_{}^{\text{}} \) & The set of tasks available within the decision window that meet theis deadlines & (\ref{eq:q}) (\ref{eq:ta_s}) (\ref{eq:ta_af})
\\
\hline

\( W_{y}^{\text{}} \) & The number of tasks within the decision window that meet their deadlines& (\ref{eq:q})  
\\
\hline
\end{tabular}
}
}
\end{table}

\subsubsection{Transmission latency}
During transmission, if multiple tasks within the RSU’s range are offloaded simultaneously, the available bandwidth is shared among them. In such cases, the bandwidth allocated to each task is proportional to its size \cite{mahsaicc}. However, if a task is transmitted individually, it can utilize the entire available bandwidth. Based on our study in \cite{mahsaicc}, sharing the bandwidth can slightly reduce transmission latency compared to using a fixed bandwidth allocation.
The set of concurrent tasks is defined as:

\begin{equation}
     \mathcal{N}
  \;=\;
  \bigl\{\,i=1,\dots,N' \;\big|\;\mathcal{T}_{i} = \mathcal{T}\bigr\}
  \,
\end{equation} $\mathcal{T}_{i}$ denotes the offloading time when task \(i\) becomes ready for transmission, while $\mathcal{T}_{}$ represents the offloading time at which all tasks in \(\mathcal{N}\) are ready to be offloaded. Additionally, \(N'\) represents the total number of tasks in set \(\mathcal{N}\).
The bandwidth assigned to task \(i\) is calculated using (\ref{eq:B}). This bandwidth is further used to calculate the transmission rate, \(r_i\), which is needed to determine the transmission time.

\begin{equation}
B_{i}^{\text{}} = 
\begin{cases} 
B_{\text{max}} \cdot \frac{S_{i}^{\text{}}}{\sum_{i=1}^{N'} S_{i} }&  {N'>1}  \\ 
B_{\text{max}}  & {N'=1}
\end{cases}
    \label{eq:B}
\end{equation}where \(S_{i}\) represents the size of task \(i\). The transmission rate is defined in (\ref{eq:r}):

\begin{equation}
r_i = B_i \cdot \log_2 \left( 1 + \frac{p \cdot g}{n_0} \right)
    \label{eq:r}
\end{equation} where \(p\) indicating the transmission power, \(g\) referring to the channel gain, and \(n_0\) denoting the noise power density.

Furthermore, the expression for the transmission time, \(t_i^{\text{tr}}\), is provided in (\ref{eq:Tcm}) and constitutes the transmission latency.

\begin{equation}
t_i^{\text{tr}}  = \frac{S_i}{r_i}
    \label{eq:Tcm}
\end{equation}

The downlink transmission follows the same process as the uplink for bandwidth allocation and calculation, and in our experimental results, their performance is very similar.

As defined in (\ref{eq:end2end}), the end-to-end latency, \(L_{i}^{\text{e2e}}\), includes both computation, \(L_i^{\text{p}}\), and transmission, \(L_i^{\text{tr}}\), latencies.

\begin{equation}
L_{i}^{\text{e2e}} = L_{i}^{\text{p}} + L_{i}^{\text{tr}}
    \label{eq:end2end}
\end{equation}

The transmission latency, \(L_i^{\text{tr}}\), includes both the uplink transmission time and the downlink transmission time.

A binary variable, \( x_{ij} \), is used to represent the assignment of task \(i\) to MEC server \(j\) as shown in (\ref{eq:assign}). Task  \(i\) is assigned to MEC \(j\) if its end-to-end latency does not exceed the duration the vehicle remains within range, \(t_i^{range}\) .

\begin{equation}
x_{ij} = 
\begin{cases} 
1 & L_i^{e2e} \leq t_i^{range} \\ 
0 & \text{otherwise} 
\end{cases}
    \label{eq:assign}
\end{equation}

\subsubsection{Number of Dropped Tasks}
The calculation for the number of dropped tasks, \({D}\), is presented in (\ref{eq:drop}). This indicates the number of tasks that were not assigned to any MEC because they could not meet their deadlines, ranging from 0 to \(N\):

\begin{equation}
D
\;=\;
\frac{1}{N}
\sum_{i=1}^{N}
\Bigl(1 \;-\;\sum_{j=1}^{M}x_{ij}\Bigr)
\label{eq:drop}
\end{equation} where \(M\) represents the total number of MECs.

\subsubsection{The Optimization Problem}
Our main objective is to minimize the number of dropped tasks, followed by reducing the end-to-end latency for non-dropped tasks. The objective function is defined in (\ref{eq:obj}). 

\begin{equation}
\text{min} \left(\lambda ({\sum_{j=1}^{M} \sum_{i=1}^{N_{\text{}}} L_{i}^{\text{e2e}} \times x_{ij}) + (1 - \lambda)  {D}}\right)
    \label{eq:obj}
\end{equation} \(\lambda\) is set to 0.4, indicating greater emphasis on \(D\). This choice reflects the critical importance of task completion in our scenario, where dropped tasks represent failures of service. While a lower value of \(\lambda\) would place even more weight on avoiding task drops, we selected this value based on preliminary empirical observations balancing both objectives.

In the dynamic scenario, the first and second tasks are assigned to MEC1 and MEC2 immediately upon arrival at the RSU. After this initial assignment, we calculate the time each MEC becomes available, as demonstrated in (\ref{eq:ava}). 

\begin{equation}
t_{j}^{\text{av}} = 
{(t_{i}^{\text{sp}} + t_{i}^{\text{p}}) x_{ij}}\quad \forall j \in {M{}^{\text{}}}
\label{eq:ava}
\end{equation} This indicates that each MEC becomes available, \(t_{j}^{\text{av}}\), only after completing the task assigned to it. 

To determine which MEC to assign the new task to, we identify the earliest MEC availability time, \(t_{e}^{\text{av}}\), using:

\begin{equation}
t_{e}^{\text{av}} = \underset{j}{\mathrm{argmin}} \left( t_{j}^{\text{av}}\right)
\label{eq:ava-s}
\end{equation}

As outlined in constraint (\ref{eq:assign-con}), every task is exclusively processed by a single MEC server. 

\begin{equation}
\sum_{j=1}^{M} x_{ij} \leq 1 \quad \forall i \in {N_{}^{\text{}}}
    \label{eq:assign-con}
\end{equation}

\begin{figure*}[!hbt]
        \centering
        \includegraphics[width = 1.0\textwidth, trim=1.3cm 11cm 1.3cm 1.3cm,clip]{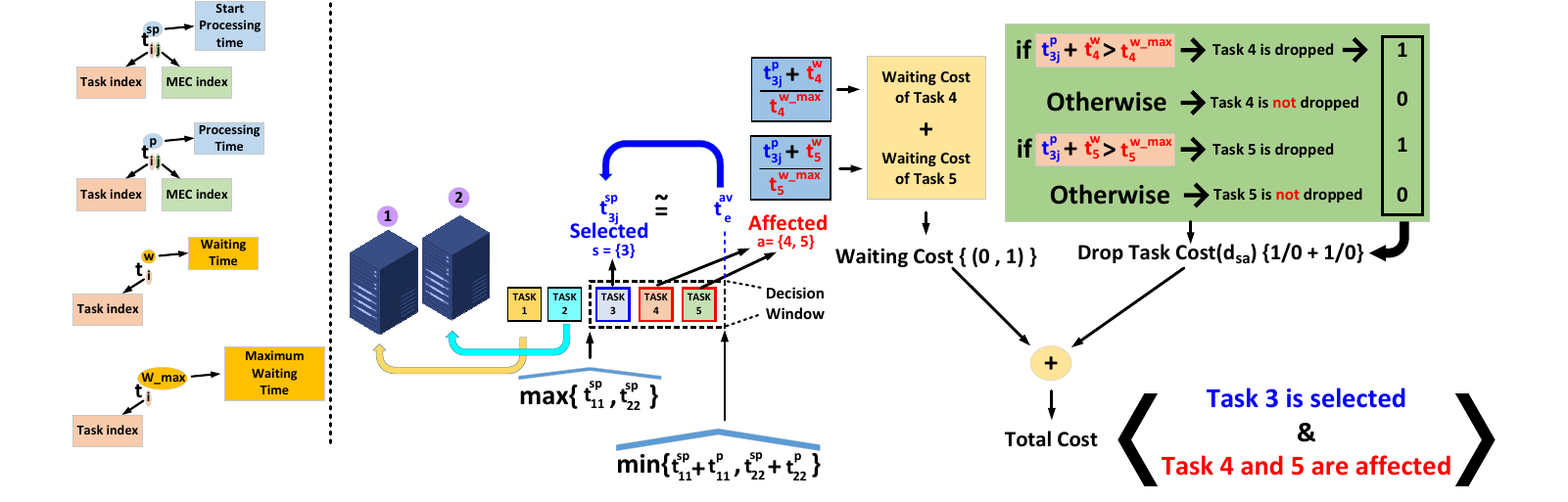}
        \caption{Illustration of a basic scenario showcasing the functionality of the On-Dyn-CDA.}
        \label{fig:cost_driven_detailed} 
\end{figure*}

\subsubsection{The Cost formula}
As shown in (\ref{eq:total_N}), \(T\) is the set of all tasks in the simulation, containing a total of \(N\) tasks. 

\begin{equation}
    T = \{t_1, t_2, t_3, \dots, t_N\}
    \label{eq:total_N}
\end{equation}

After assigning a task to a MEC server, a new set of tasks, \(S\), is formed as defined in (\ref{eq:s}). This set includes tasks that arrive before the earliest availability time, \(t_{e}^{av}\),  and have not yet been assigned to any MEC server.

\begin{equation}
    S = \{t_{i_1}, t_{i_2}, t_{i_3}, \dots, t_{i_W}\}
    \label{eq:s}
\end{equation}

This implies that the arrival time of tasks, \(t_{il}^{ar}\), in \(S\) is less than or equal to the earliest availability time, \(t_{e}^{av}\), based on (\ref{eq:tik}).

\begin{equation}
    t_{il}^{ar} \leq t_{e}^{av}
    \label{eq:tik}
\end{equation}

The set \(Q_y\) is a subset of tasks from \(S\) that satisfy the condition of being able to meet their deadlines, as determined in (\ref{eq:q}). This means that only tasks that arrive before \(t_{e}^{av}\) and can meet their deadlines are included in the \(Q_y\) set, which represents our decision window. Tasks that cannot meet their deadlines are excluded from this window, as they are not eligible for assignment:

\begin{equation}
\begin{aligned}
Q_y = \bigl\{t_{y_l} \in S \;\big|\; &\; 1 \leq l \leq W_y, \; t_{y_l}^{ar} \leq t_e^{\text{av}}, \\
&\; t_{y_l}^w \leq t_{y_l}^{range} - t_{y_l}^p - t_{y_l}^\text{cm} \bigr\} \hspace*{1.5em}
\end{aligned}
\label{eq:q}
\end{equation} where \(t_{y_l}\) denotes the task at position \(l\) in window \(y\).
As demonstrated in \figurename\hspace{0.1pt}~\ref{fig:cost_driven_detailed}, if there are \(W_y\) tasks in \(Q_y\) (with \(W_y = 3\) illustrated in \figurename \hspace{0.1pt}~\ref{fig:cost_driven_detailed}), there are \(W_y\) possible tasks that can be selected to be assigned first to a MEC server once it becomes available. In other words, each task within the decision window that belongs to set \(Q_y\) can be selected to be sent to the MEC server with the earliest availability.

Our goal is to choose the task that minimizes overall latency and results in fewer tasks being dropped. As specified in (\ref{eq:ta_s}), each task within the decision window's \(Q_y\) set is evaluated as the selected task, \(t_s\), while the remaining tasks are treated as affected tasks and are included in the set \(Q_{ya}\), as outlined in (\ref{eq:ta_af}). 

\begin{equation}
    t_s \in Q_y
    \label{eq:ta_s}
\end{equation}

\begin{equation}
    Q_{ya} = Q_y \setminus \{t_s\}
    \label{eq:ta_af}
\end{equation}

This terminology is used because we are deciding which task should be assigned first (the selected task), while the affected tasks are those influenced by the selected task. Specifically, the selected task can influence the waiting time of the affected tasks, and a poor choice for this task could result in significant delays for the others. Our goal is to identify the selected task that results in the fewest affected tasks being dropped while keeping the waiting time for the remaining, non-dropped affected tasks as low as possible. After assigning the optimal selected task to the MEC, we calculate the updated availability times of the MECs to determine the new earliest MEC availability time. A new decision window is created, spanning from the start processing time of the assigned task (previous earliest availability time) to the newly determined earliest availability time. This window includes the non-dropped affected tasks as well as any tasks that arrive before the new earliest availability time. We repeat the process to determine the next optimal task for assignment, with only one task being assigned to a MEC at a time. Maximum waiting time, \({t_{i}^{\text{w\_max}}}\), as determined in (\ref{eq:max_wait}), represents the longest period that task \(i\) can wait before being processed while still meeting its deadline. The task is dropped if it exceeds this period.

\begin{equation}
{t_{i}^{\text{w\_max}}} = 
t_{i}^{\text{d}} - t_{i}^{\text{p}} - t_{i}^{\text{ar}}\hspace{8pt} \forall i \in Q_y\\
\label{eq:max_wait}
\end{equation}

According to (\ref{eq:sp_s}), the selected task's start processing time, \({t_{s}^{\text{sp}}}\), is equal to the earliest availability time of the MEC server, \({t_{e}^{\text{av}}}\), added to the algorithm's execution time, \(\epsilon_c\). The execution time is high in the PSO algorithm, whereas it is significantly lower in On-Dyn-CDA. 

\begin{equation}
    {t_{s}^{\text{sp}}} = {t_{e}^{\text{av}}} + \epsilon_c
    \label{eq:sp_s}
\end{equation}

The waiting time in window \(y\) for an affected task, \({t_{a}^{\text{w}}}\), is defined as the duration between an affected task's arrival time and the earliest availability time at the moment when a specific selected task is being evaluated. In other words, this waiting time is only used to determine the optimal selected task and does not represent the final waiting time. This implies that when evaluating a particular task as the selected task, we determine the waiting time it imposes on the affected tasks. This method helps assess how choosing a suboptimal selected task could lead to longer waiting times for subsequent tasks. As defined in (\ref{eq:d-cost}), \(d_{sa}\) is a binary variable that indicates whether an affected task is dropped as a result of selecting a specific task to be assigned to the MEC. If the sum of the selected task's processing time, \({t_{sj}^{\text{p}}}\), and the waiting time in window \(y\), \({t_{a}^{\text{w}}}\), of the affected task is less than or equal to the maximum waiting time, \({t_{a}^{\text{w\_max}}}\), the affected task will not be dropped. Otherwise, it is dropped due to the impact of the selected task.

\begin{equation}
    d_{sa} =
    \begin{cases}
        0 & \text{if } {t_{s}^{\text{p}}} + {t_{a}^{\text{w}}} \leq {t_{a}^{\text{w\_max}}} \\
        1 & \text{otherwise}
    \end{cases}
    \label{eq:d-cost}
\end{equation}

To dynamically approximate the global objective defined in (\ref{eq:obj}), which jointly minimizes end-to-end latency and the number of dropped tasks across the entire task set, the On-Dyn-CDA algorithm employs a local heuristic objective given in (\ref{eq:final-cost-min}).

\begin{equation}
    \operatorname*{arg\,min}_s \left\{ \sum_{\substack{a=1 \\ a \neq s}}^{W_y} \left( \frac{{t_{s}^{\text{p}}} + {t_{a}^{\text{w}}}}{{t_{a}^{\text{w\_max}}}} \cdot (1 - d_{sa}) + d_{sa} \right) \right\}
    \label{eq:final-cost-min}
\end{equation} This equation serves as a tractable, local approximation that guides dynamic decision-making at each step of the task assignment process. Specifically, (\ref{eq:final-cost-min}) is used to evaluate the impact of selecting a particular task on the waiting time and drop likelihood of other tasks in the decision window.

\vspace{1mm}
\noindent\rule{\linewidth}{0.4pt}

\noindent\textbf{Algorithm 1:} On-Dyn-CDA: Online Dynamic Cost-Driven Assignment

\vspace{-1mm}
\noindent\rule{\linewidth}{0.4pt}

\begin{algorithmic}[1]
\State \textbf{Initialization:}
\State Assign first two tasks to MECs and compute $t_j^{av}$ for each.

\While{unassigned tasks remain}
    \State Compute $t_e^{av} = \arg\min_j(t_j^{av})$
    \State $Q_y \gets \{t_i \mid t_i^{ar} \leq t_e^{av}, \text{meets deadline} \}$
    
    \For{each $t_s \in Q_y$}
        \State $cost_s \gets 0$, $Q_{ya} \gets Q_y \setminus \{t_s\}$
        \For{each $t_a \in Q_{ya}$}
            \State Compute $t_a^w$, $t_a^{w\_max}$
            \If{$t_s^p + t_a^w \leq t_a^{w\_max}$}
                \State $cost_s \mathrel{+}= \frac{t_s^p + t_a^w}{t_a^{w\_max}}$
            \Else
                \State $cost_s \mathrel{+}= 1$
            \EndIf
        \EndFor
    \EndFor

    \State Select $t_{\text{opt}} = \arg\min(cost_s)$
    \State Assign $t_{\text{opt}}$, update $t_j^{av}$

    \State Update window with remaining and new tasks
\EndWhile
\end{algorithmic}

\noindent\rule{\linewidth}{0.4pt}

By iteratively applying this heuristic in a greedy fashion, always selecting the task that minimizes this local cost, the On-Dyn-CDA algorithm approximates good solutions to the global problem without incurring its computational overhead.
This means that, to determine the best selected task for assignment to the MEC, we seek the one that minimizes the number of dropped tasks and reduces the waiting impact on those that are not dropped. If \(d_{sa}\) is 1, there is no need to check the impact of the selected task on the waiting time of the affected task. However, if it is 0, this impact must be evaluated as the ratio of the sum of the selected task's processing time, \({t_{sj}^{\text{p}}}\), and the affected task's waiting time in window \(y\), \({{t_{a}^{\text{w}}}}\), to the maximum waiting time, \({{t_{a}^{\text{w\_max}}}}\). This ratio can be referred to as the waiting cost since it measures the relative delay experienced by an affected task. The value of \(d_{sa}\) is limited to either 0 or 1, whereas the waiting cost ranges between 0 and 1.

The formulation in (\ref{eq:final-cost-min}) is effective because it balances two key aspects of task scheduling: feasibility (whether a task is dropped) and timeliness (how close a task is to exceeding its deadline). The binary term \(d_{sa}\) penalizes cases where the affected task \(a\) would be dropped due to the selection of task \(s\), directly addressing the feasibility constraint. For non-dropped tasks (\(d_{sa}=0\)), the cost is expressed as a normalized ratio which measures the portion of the affected task's maximum waiting time that is consumed when task \(s\) is selected. A lower ratio implies that the affected task remains well within its allowable waiting time, minimizing its risk of deadline violation. In this way, the expression effectively quantifies the waiting cost experienced by each affected task. By combining the drop penalty, \(d_{sa}\), and the relative waiting cost, the objective in (\ref{eq:final-cost-min}) prioritizes task assignments that not only avoid drops but also preserve temporal flexibility for future scheduling decisions. This makes it a computationally efficient yet practical approximation of the global objective in (\ref{eq:obj}).

\subsection{Algorithm}
\notemahsanew{Algorithm 1 outlines the On-Dyn-CDA method for real-time task offloading in MEC-enabled environments. The algorithm works by repeatedly selecting the most efficient task from the decision window to assign to the MEC server that will become available next. For each task in the decision window, the algorithm estimates a cost associated with assigning it next. This cost is determined by examining how the assignment would affect other tasks that are also waiting to be scheduled. If a task’s total time in the system, which is the sum of its waiting time and processing time, remains within its allowed deadline, the algorithm adds a proportional cost. This cost is computed as the ratio of the task’s total time to its maximum tolerable waiting time. It reflects how close the task is to missing its deadline. The closer the task is to this limit, the higher the cost, which indicates an increased scheduling risk.
If simulating the assignment causes any waiting task to exceed its allowed deadline, a fixed penalty cost is added instead. This indicates a more severe outcome, such as a deadline violation, which the algorithm strongly prefers to avoid. By using both proportional and penalty-based costs, the algorithm is able to distinguish between low-risk and high-risk assignments.}


\section{Performance Evaluation}
\subsection{Experimental Setup}
In this study, we simulate a highway environment using the SUMO platform, where an RSU facilitates computational offloading for vehicles traveling along the highway. Vehicle mobility is dynamically controlled using TraCI, allowing real-time interaction with the simulation to reflect realistic traffic behavior and offloading conditions. The RSU is equipped with two MEC servers to process incoming tasks generated by vehicles within its coverage area. Each task has a time-sensitive deadline, determined by the Euclidean distance between the vehicle and the RSU at the time of offloading. The simulation examines three traffic density scenarios to evaluate the performance of the MEC servers under varying workloads. The maximum number of tasks is limited to 200, as reducing task drops become increasingly difficult beyond this point due to the short deadlines in our scenario. 

Each vehicle generates a single object detection task, with task generation following a Poisson distribution to capture the variability typical of real-world vehicular offloading. The choice of the Poisson distribution is motivated by its simplicity and ease of implementation. Tasks can be generated either while a vehicle is within the RSU’s range or before entering it, creating a more realistic simulation. \notemahsanew{The size of each task is determined by the resolution of the image being processed for object detection, as inspired by \cite{execution}, \cite{tasksize2}. To model different computational loads, we simulate varying image resolutions, with the data size of each image calculated based on a color depth of 24 bits per pixel, following standard RGB encoding \cite{tasksize1}.} The parameters have been set based on \cite{para1}, \cite{para2}, as presented in Table III.

\begin{table}[!t]
\centering 
\renewcommand{\arraystretch}{1.2} 
\caption{Simulation Parameters} 
\begin{tabular}{|l|l|}
\hline
\textbf{Parameters} & \textbf{Value}   \\   \hline
The number of vehicles & 50, 100, 200\\
\hline
The number of tasks per vehicle & 1 \\   \hline
The number of MEC servers & 2   \\   \hline
The number of CPUs per MEC server & 1   \\   \hline
Task size & 2160, 3840, 6000, 8640 kb\\ \hline

Max bandwidth & 20 Mhz   \\   \hline
Noise power & 100 dbm   \\   \hline
Transmit power & 200 mW \\   \hline

\end{tabular}
\label{tab:parameters} 
\end{table}

\notemahsanew{The system operates with a maximum bandwidth of 20 MHz \cite{bandwidthsize}; however, accounting for a 4.6\% guard band, the effective usable bandwidth is 19.08 MHz \cite{guardband}.} Simulations are performed on a machine with a Core i7 CPU, GeForce MX150 GPU, and 8GB of RAM. The parameters used in the PSO algorithms \cite{parisa2} are detailed in Table IV. These parameters are selected based on empirical observations from multiple experiments.

\begin{table}[!t]
\centering 
\renewcommand{\arraystretch}{1.2} 
\caption{PSO Parameters}
\label{table:ga_parameters}
\begin{tabular}{|l|l|}
\hline
\textbf{Parameters} & \textbf{Value} \\
\hline
Swarm Size & 50 \\
\hline
Maximum number of iterations & 100 \\
\hline
Cognitive Coefficient & 1.49 \\
\hline
Social Coefficient & 1.49 \\
\hline

\end{tabular}
\label{tab:PSO}
\end{table}

\subsection{Numerical Results}

\begin{table*}[h!]
 \caption{Overall performance of all methods}
    \centering
    \scalebox{0.87}{
    \renewcommand{\arraystretch}{1.5}
    \begin{tabular}{|c|c|c|c|c|c|c|c|c|c|c|}
        \hline
        \multirow{2}{*}{\textbf{Algorithm}} & \multicolumn{3}{|c|}{\textbf{Total end-end Latency (s)}} & \multicolumn{3}{|c|}{\textbf{Total waiting time (s)}} & \multicolumn{3}{|c|}{\textbf{Algorithm Execution Time (s)}} \\ 
        \cline{2-10}
        & \textbf{50 Vehicles} & \textbf{100 Vehicles} & \textbf{200 Vehicles} & \textbf{50 Vehicles} & \textbf{100 Vehicles} & \textbf{200 Vehicles} & \textbf{50 Vehicles} & \textbf{100 Vehicles} & \textbf{200 Vehicles} \\
        \hline
        FCFS & 413.44 & 789.02 & 1104.78 & 278.32 & 637.25 & 955.01 & 0.0052 & 0.0078 & 0.02 \\
        \hline
        SDF & 430.03 & 792.12 & 1105.18 & 292.57 & 642.05 & 956.32 & 0.0075 & 0.0117 & 0.0313 \\
        \hline
        Off-Sta-PSO & 392.49 & 675.75 & 861.32 & 270.21 & 453.61 & 600.84 & 210.61 & 800.02 & 1300.02 \\
        \hline
        On-Sta-PSO & 483.10 & 881.28 & 1253.80 & 339.42 & 724.65 & 1112.91 & 213.42 & 805.10 & 1308.24 \\
        \hline
        On-Dyn-PSO & 460.12 & 827.12 & 1175.78 & 314.93 & 673.7 & 1013.33 & 217.54 & 821.30 & 1330.05 \\
        \hline
        \textbf{On-Dyn-CDA} & \textbf{396} & \textbf{686.2} & \textbf{872.62} & \textbf{278.11} & \textbf{471.12} & \textbf{632.12} & \textbf{0.02} & \textbf{0.03} & \textbf{0.05} \\
        \hline

    \end{tabular}
    }
   
    \label{tab:full_width_table}
\end{table*}

FCFS and SDF are employed as deterministic methods to demonstrate how simple approaches perform in addressing a complex task offloading problem. PSO is applied in various configurations, including different simulation types (online and offline) and task processing strategies (static and dynamic), to evaluate the performance of a well-known optimization algorithm in diverse environments. Finally, we introduce a novel algorithm, On-Dyn-CDA, designed to overcome the inefficiency of traditional optimization algorithms in dynamic scenarios. This algorithm achieves results in terms of the number of dropped tasks and end-to-end latency that are close to the theoretical upper bound but with significantly reduced algorithm execution time. This highlights that On-Dyn-CDA not only meets the primary objectives but also processes a large number of tasks quickly, making it highly suitable for real-time task offloading.

\subsubsection{Convergence}
\begin{figure}[!hbt]
        \centering
        \includegraphics[width=0.45\textwidth, trim=0cm 0.0cm 0cm 0cm, clip]{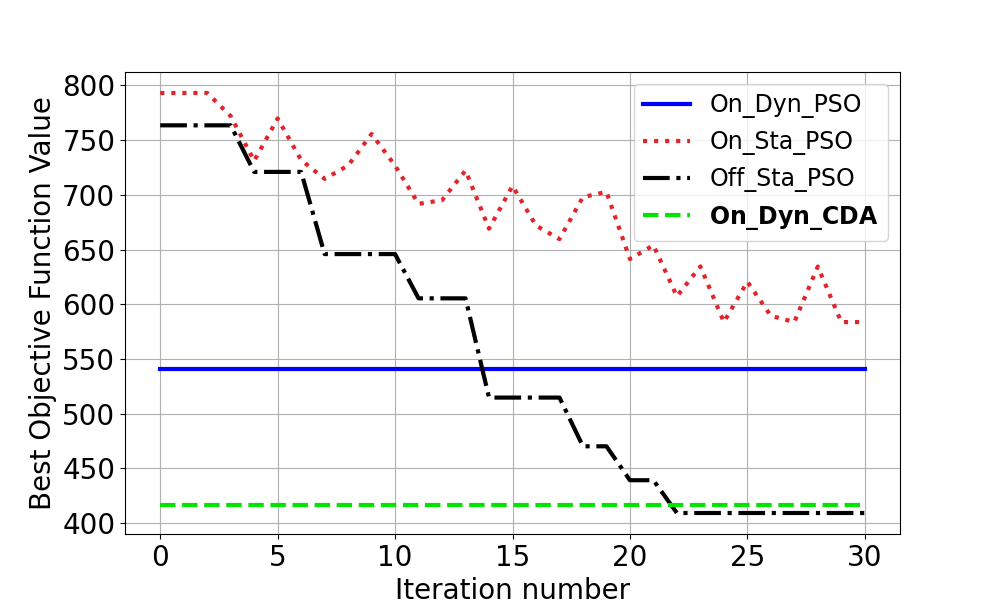}
        \caption{Comparison of Convergence Plots for different algorithms with 200 tasks}
        \label{fig:convergences} \vspace{-2mm}
\end{figure}

In \figurename \hspace{0.1pt}\ref{fig:convergences}, we plot the objective function values over the iterations for 200 tasks. The convergence pattern of On-Sta-PSO, as shown in the figure, demonstrates its difficulty in finding the global optimum. This is primarily due to the significant waiting time introduced in this approach, which makes it challenging for the algorithm to explore the solution space effectively. Since processing does not begin until all tasks have arrived and PSO has been executed, the extended delays lead to inefficiencies, causing the algorithm to struggle in balancing task assignments and minimizing the objective function. These constraints explain the slower convergence and the tendency to get stuck in suboptimal solutions. Off-Sta-PSO demonstrates good convergence as it does not account for the significant waiting times. It successfully finds the global optimum, achieving the best possible value. For On-Dyn-PSO, which is executed multiple times on different batches of tasks, the objective function value is calculated at the end based on the final total order. The same approach is applied for On-Dyn-CDA. The results indicate that the objective value of On-Dyn-CDA is better than that of On-Dyn-PSO, primarily because of the significant execution time required by PSO.

\subsubsection{Dropped Tasks}
\begin{figure}[!hbt]
        \centering
        \includegraphics[width=0.45\textwidth, trim=0cm 0.0cm 0cm 0cm, clip]{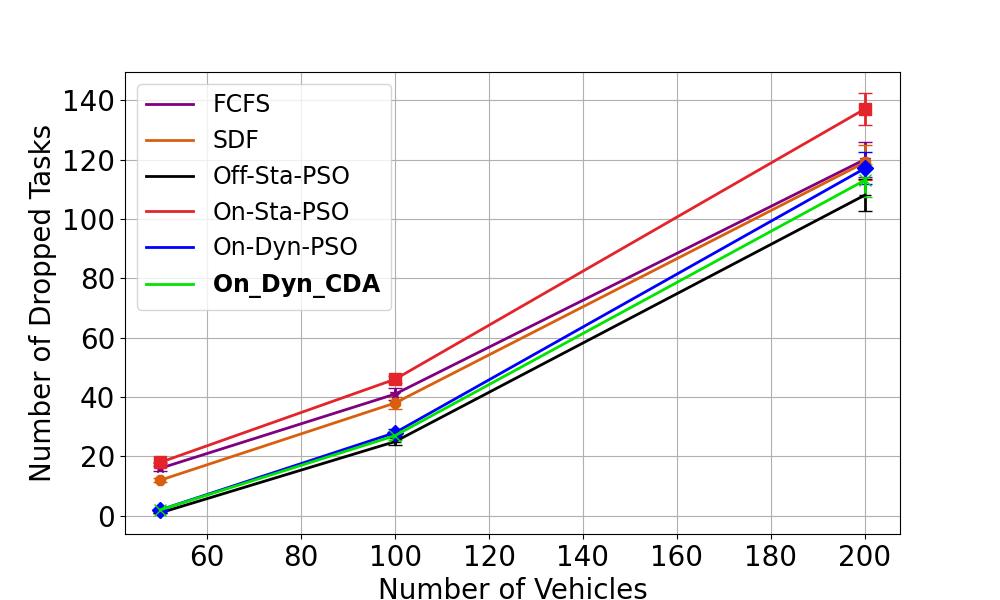}
        \caption{Number of Dropped Tasks for different numbers of vehicles across various algorithms}
        \label{fig:drop} \vspace{-2mm}
\end{figure}

\figurename \hspace{0.1pt}\ref{fig:drop} shows the number of dropped tasks for different vehicle counts, including confidence intervals, across various algorithms over 10 runs. This indicates that as the number of tasks increases, the number of dropped tasks also rises across all algorithms due to longer waiting times. Off-Sta-PSO achieves the best performance, while On-Sta-PSO performs the worst. This is due to the fact that Off-Sta-PSO assumes an ideal scenario, where all tasks are scheduled simultaneously without accounting for real-time waiting delays, whereas On-Sta-PSO reflects a worst-case scenario using static PSO under real-time conditions, accounting for both waiting and execution times. This leads to more tasks missing their deadlines. On-Dyn-CDA delivers results very close to the theoretical upper bound of Off-Sta-PSO, demonstrating strong performance. On-Dyn-PSO performs worse than On-Dyn-CDA but still achieves better results than FCFS and SDF. Despite the dynamic nature of On-Dyn-PSO, its performance is affected by the considerable execution time required by PSO to process different batches of tasks. SDF performs better than FCFS by reducing the number of dropped tasks. However, as the number of tasks increases, the results of SDF and FCFS converge. This happens because the complexity of the task processing environment increases with a larger number of computationally intensive tasks, making SDF less efficient in maintaining its advantage over FCFS.

\begin{figure}[!hbt]
        \centering
        \includegraphics[width=0.45\textwidth, trim=0cm 0.0cm 0cm 0cm, clip]{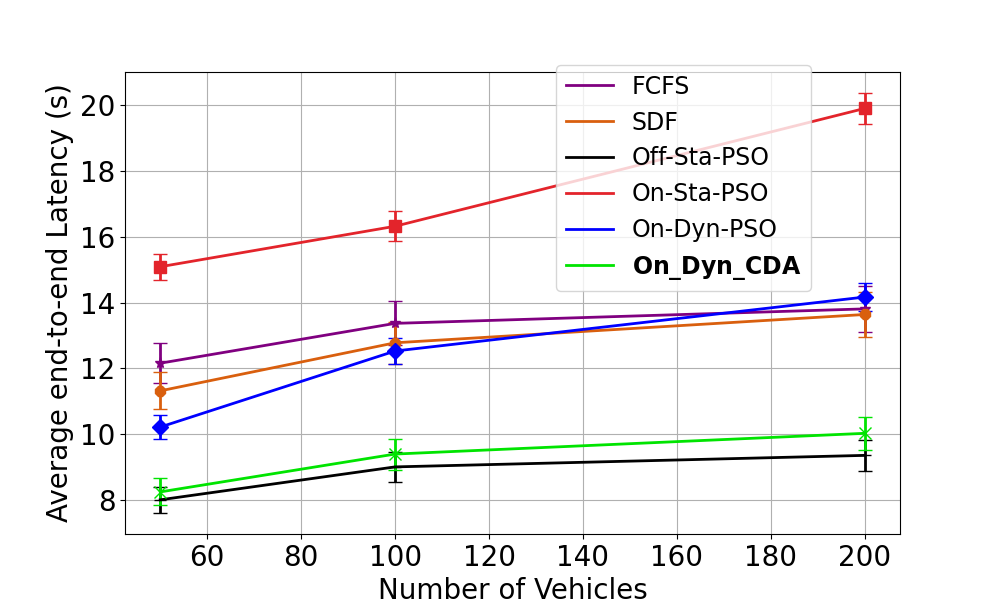}
        \caption{Average end-to-end latency for different numbers of vehicles
        across various algorithms}
        \label{fig:latency} \vspace{-2mm}
\end{figure}

\begin{figure}[!hbt]
        \centering
        \includegraphics[width=0.45\textwidth, trim=0cm 0.0cm 0cm 0cm, clip]{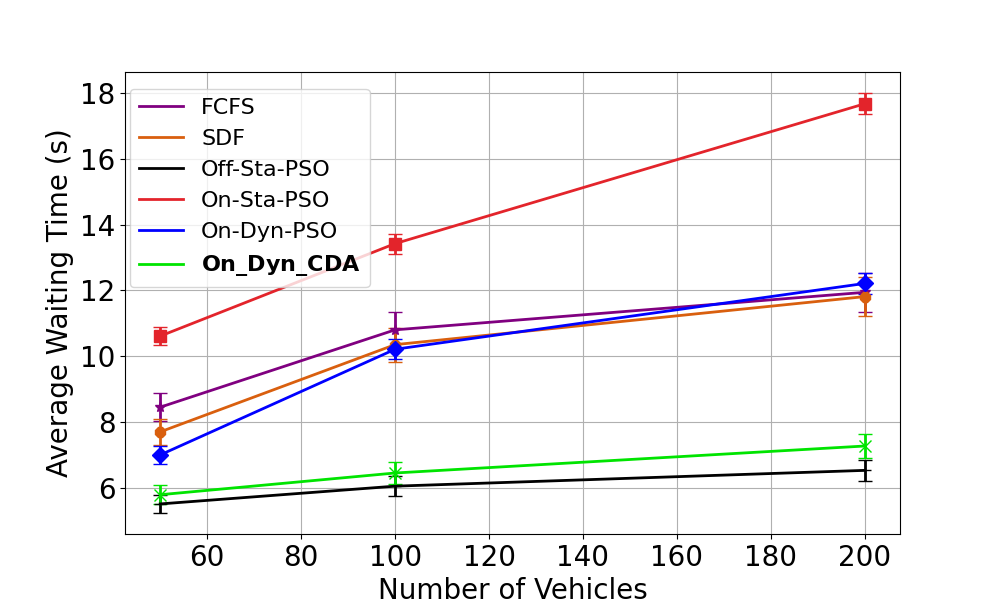}
        \caption{Average waiting time for different numbers of vehicles
        across various algorithms}
        \label{fig:waiting} \vspace{-2mm}
\end{figure}

\subsubsection{Computation Latency}

\begin{figure}[!hbt]
        \centering
        \includegraphics[width=0.45\textwidth, trim=0cm 0.0cm 0cm 0cm, clip]{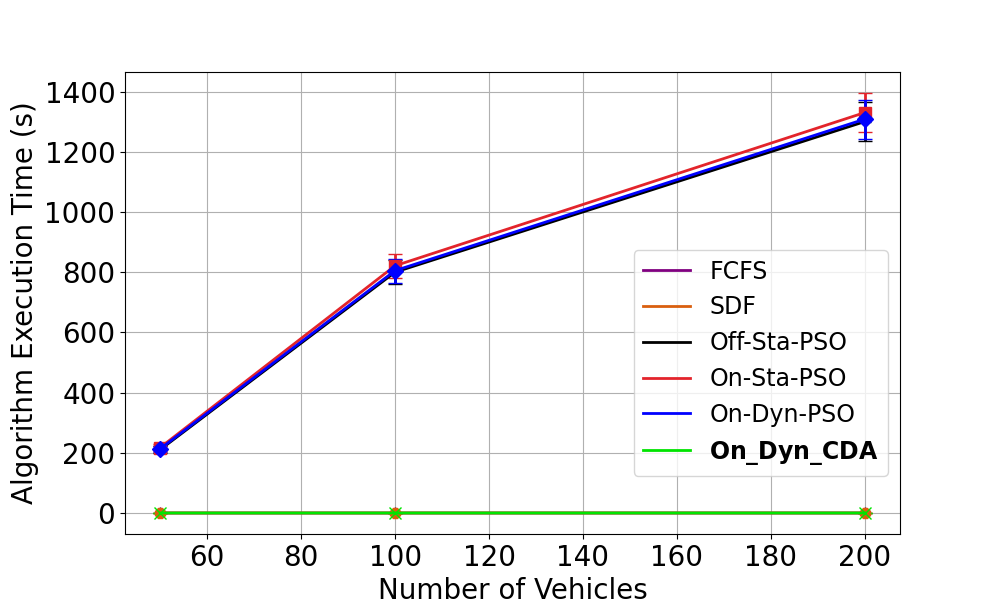}
        \caption{Algorithm execution time for different number of users under different algorithms}
        \label{fig:all_simulation_time} \vspace{-2mm}
\end{figure}

The comparison of average end-to-end latency is presented in \figurename\hspace{0.1pt}\ref{fig:latency} for different vehicle counts, including confidence intervals, across various algorithms over 10 runs. Among the evaluated approaches, Off-Sta-PSO and On-Dyn-CDA demonstrate the best performance, with Off-Sta-PSO representing the theoretical optimum. On-Dyn-PSO initially achieves a lower average end-to-end latency compared to FCFS and SDF. However, as the number of tasks increases, the execution time of On-Dyn-PSO plays a major role in increasing the overall latency, leading to a decline in its performance compared to FCFS and SDF. Notably, On-Sta-PSO consistently performs the worst, exhibiting the highest average end-to-end latency across all scenarios. \figurename \hspace{0.1pt}\ref{fig:waiting} presents the average waiting time for different vehicle counts, including confidence intervals, across various algorithms over 10 runs. As shown, the average waiting time results are similar to those observed for the average end-to-end latency, since waiting time serves as the primary contributing factor to end-to-end latency. In \figurename \hspace{0.1pt}\ref{fig:all_simulation_time} the execution time for all algorithms is displayed, indicating the overall duration required for each algorithm to process all tasks.
As observed, FCFS, SDF, and On-Dyn-CDA exhibit the shortest algorithm execution time, whereas the PSO-based models require significantly more time. As expected, On-Sta-PSO demonstrates the longest execution time, followed by On-Dyn-PSO, with Off-Sta-PSO performing better but still requiring more time compared to the other approaches. On-Dyn-PSO requires more execution time than Off-Sta-PSO because it performs real-time optimizations, continuously updating task assignments and server selections based on changing system conditions. In contrast, Off-Sta-PSO precomputes schedules offline under static conditions, avoiding the need for continuous updates and reducing computational overhead.

\begin{figure}[!hbt]
        \centering
        \includegraphics[width=0.45\textwidth, trim=0cm 0.0cm 0cm 0cm, clip]{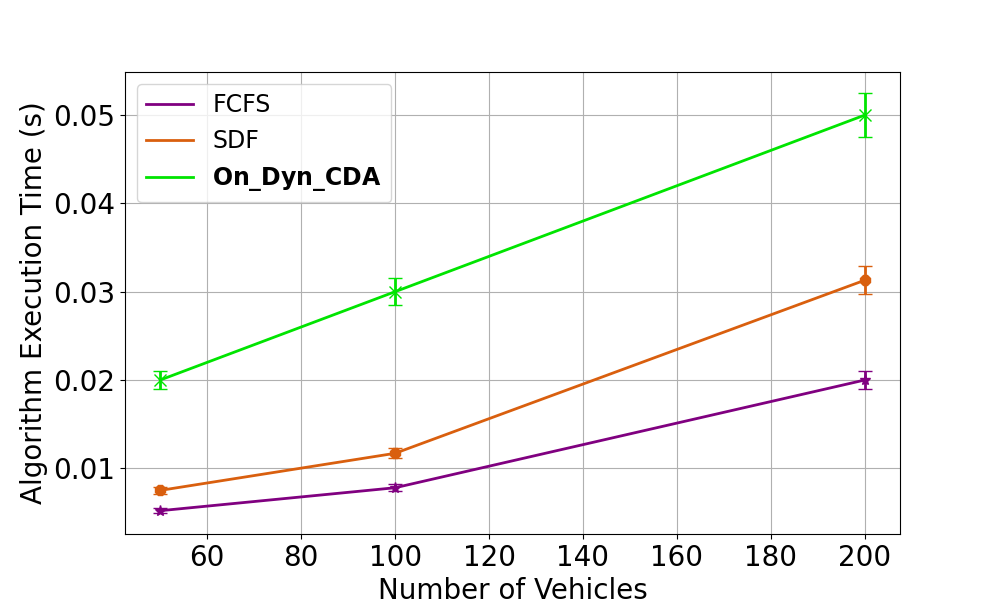}
        \caption{Algorithm execution time for different numbers of vehicles for FCFS, SDF, and On-Dyn-CDA}
        \label{fig:online_simulation_time} \vspace{-2mm}
\end{figure}

For better comparison, \figurename \hspace{0.1pt}\ref{fig:online_simulation_time} presents the execution time of only the non-PSO algorithms. As illustrated, FCFS has the shortest execution time, followed by SDF, and finally On-Dyn-CDA. Although FCFS and SDF offer shorter execution times than On-Dyn-CDA, they perform poorly in reducing latency and task loss. Table V presents the numerical results for total end-to-end Latency, total waiting time, and algorithm execution time. As shown in \figurename \hspace{0.1pt}\ref{fig:all_simulation_time} and the numerical values in Table V, the execution time for On-Dyn-CDA does not increase significantly with the number of vehicles, unlike PSO-based algorithms. This demonstrates the scalability of the On-Dyn-CDA algorithm.

\section{Conclusion}
Our comparative study of task offloading algorithms highlights the trade-offs between simplicity, optimization, and real-time performance in dynamic computational environments. FCFS and SDF, while straightforward, struggle to maintain efficiency as task complexity increases. PSO, when applied in offline and online configurations, demonstrates varying strengths and limitations. The offline static approach achieves the theoretical optimum, offering the best possible performance in terms of dropped tasks and latency. However, it lacks adaptability to dynamic scenarios, making it impractical for real-world applications. On-Dyn-PSO, while dynamic, suffers from a large execution time that limits its effectiveness in fast-changing environments. The proposed On-Dyn-CDA algorithm addresses this gap by outperforming On-Dyn-PSO with a 3.42\% reduction in dropped tasks and a 29.22\% decrease in average latency, based on experiments with 200 tasks. It demonstrates a faster execution time compared to PSO in the dynamic approach and achieves performance closest to the theoretical upper bound represented by Off-Sta-PSO, compared to baseline models. Its superior adaptability and simplicity demonstrate that it is an efficient and scalable solution for addressing dynamic computational demands. In future work, we plan to focus on energy optimization and the incorporation of task dependencies to further improve the practicality and efficiency of the task offloading process. Additionally, we will explore the potential of ML models, such as RL, and compare their performance with On-Dyn-CDA to identify optimal solutions for complex and dynamic offloading environments.

\section*{Acknowledgment}
This work was supported in part by funding from the Innovation for Defence Excellence and Security (IDEaS) program from the Department of National Defence (DND), and in part by Natural Sciences and Engineering Research Council of Canada (NSERC) CREATE TRAVERSAL Program.

\normalem
\bibliographystyle{IEEEtran}

\begin{thebibliography}{10}
\providecommand{\url}[1]{#1}
\csname url@samestyle\endcsname
\providecommand{\newblock}{\relax}
\providecommand{\bibinfo}[2]{#2}
\providecommand{\BIBentrySTDinterwordspacing}{\spaceskip=0pt\relax}
\providecommand{\BIBentryALTinterwordstretchfactor}{4}
\providecommand{\BIBentryALTinterwordspacing}{\spaceskip=\fontdimen2\font plus
\BIBentryALTinterwordstretchfactor\fontdimen3\font minus \fontdimen4\font\relax}
\providecommand{\BIBforeignlanguage}[2]{{%
\expandafter\ifx\csname l@#1\endcsname\relax
\typeout{** WARNING: IEEEtran.bst: No hyphenation pattern has been}%
\typeout{** loaded for the language `#1'. Using the pattern for}%
\typeout{** the default language instead.}%
\else
\language=\csname l@#1\endcsname
\fi
#2}}
\providecommand{\BIBdecl}{\relax}
\BIBdecl

\bibitem{iot6}
Y.~Sun, Z.~Wu, K.~Meng, and Y.~Zheng, ``Vehicular task offloading and job scheduling method based on cloud-edge computing,'' \emph{IEEE Transactions on Intelligent Transportation Systems}, vol.~24, no.~12, pp. 14\,651--14\,662, 2023.

\bibitem{iot1}
Z.~Wang, G.~Sun, H.~Su, H.~Yu, B.~Lei, and M.~Guizani, ``Low-latency scheduling approach for dependent tasks in mec-enabled 5g vehicular networks,'' \emph{IEEE Internet of Things Journal}, vol.~11, no.~4, pp. 6278--6289, 2024.

\bibitem{iot2}
M.~A. Mirza, Y.~Junsheng, S.~Raza, M.~Ahmed, M.~Asif, A.~Irshad, and N.~Kumar, ``Mcla task offloading framework for 5g-nr-v2x-based heterogeneous vecns,'' \emph{IEEE Transactions on Intelligent Transportation Systems}, vol.~24, no.~12, pp. 14\,329--14\,346, 2023.

\bibitem{iot3}
L.~Li and P.~Fan, ``Latency and task loss probability for noma assisted mec in mobility-aware vehicular networks,'' \emph{IEEE Transactions on Vehicular Technology}, vol.~72, no.~5, pp. 6891--6895, 2023.

\bibitem{iot5}
J.~Du, Y.~Sun, N.~Zhang, Z.~Xiong, A.~Sun, and Z.~Ding, ``Cost-effective task offloading in noma-enabled vehicular mobile edge computing,'' \emph{IEEE Systems Journal}, vol.~17, no.~1, pp. 928--939, 2023.

\bibitem{iot4}
\BIBentryALTinterwordspacing
X.~He, Y.~Cen, Y.~Liao, X.~Chen, and C.~Yang, ``Optimal task offloading strategy for vehicular networks in mixed coverage scenarios,'' \emph{Applied Sciences}, vol.~14, no.~23, 2024.
\BIBentrySTDinterwordspacing

\bibitem{iot7}
Y.~Lu, D.~Han, X.~Wang, and Q.~Gao, ``Distributed task offloading for large-scale vec systems: A multi-agent deep reinforcement learning method,'' in \emph{2022 14th International Conference on Communication Software and Networks (ICCSN)}, 2022, pp. 161--165.

\bibitem{iot8}
B.~Wang, L.~Liu, and J.~Wang, ``Multi-agent deep reinforcement learning for task offloading in vehicle edge computing,'' in \emph{2023 IEEE International Symposium on Broadband Multimedia Systems and Broadcasting (BMSB)}, 2023, pp. 1--6.

\bibitem{iot9}
P.~Dai, Y.~Huang, K.~Hu, X.~Wu, H.~Xing, and Z.~Yu, ``Meta reinforcement learning for multi-task offloading in vehicular edge computing,'' \emph{IEEE Transactions on Mobile Computing}, vol.~23, no.~3, pp. 2123--2138, 2024.

\bibitem{iot10}
B.~Zhang, F.~Xiao, and L.~Wu, ``Offline reinforcement learning for asynchronous task offloading in mobile edge computing,'' \emph{IEEE Transactions on Network and Service Management}, vol.~21, no.~1, pp. 939--952, 2024.

\bibitem{IoT-ieee6}
A.~Naouri, H.~Wu, N.~A. Nouri, S.~Dhelim, and H.~Ning, ``A novel framework for mobile-edge computing by optimizing task offloading,'' \emph{IEEE Internet of Things Journal}, vol.~8, no.~16, pp. 13\,065--13\,076, 2021.

\bibitem{intro1}
M.~A. Ferrag, O.~Friha, B.~Kantarci, N.~Tihanyi, L.~Cordeiro, M.~Debbah, D.~Hamouda, M.~Al-Hawawreh, and K.-K.~R. Choo, ``Edge learning for 6g-enabled internet of things: A comprehensive survey of vulnerabilities, datasets, and defenses,'' \emph{IEEE Communications Surveys \& Tutorials}, vol.~25, no.~4, pp. 2654--2713, 2023.

\bibitem{intro2}
B.~Radouane, G.~Lyamine, K.~Ahmed, and B.~Kamel, ``Scalable mobile computing: From cloud computing to mobile edge computing,'' in \emph{2022 5th International Conference on Networking, Information Systems and Security: Envisage Intelligent Systems in 5g//6G-based Interconnected Digital Worlds (NISS)}, 2022, pp. 1--6.

\bibitem{parisaicc}
\BIBentryALTinterwordspacing
P.~F. Moshiri, M.~Simsek, and B.~Kantarci, ``Partitioned task offloading for low-latency and reliable task completion in 5g mec,'' 2025.
\BIBentrySTDinterwordspacing

\bibitem{intro3}
J.~Lee and W.~Na, ``A survey on vehicular edge computing architectures,'' in \emph{2022 13th International Conference on Information and Communication Technology Convergence (ICTC)}, 2022, pp. 2198--2200.

\bibitem{AVCIL2024100773}
\BIBentryALTinterwordspacing
M.~N. Avcil, M.~Soyturk, and B.~Kantarci, ``Fair and efficient resource allocation via vehicle-edge cooperation in 5g-v2x networks,'' \emph{Vehicular Communications}, p. 100773, 2024.
\BIBentrySTDinterwordspacing

\bibitem{IoT-ieee8}
T.~H. Binh, D.~B. Son, H.~Vo, B.~M. Nguyen, and H.~T.~T. Binh, ``Reinforcement learning for optimizing delay-sensitive task offloading in vehicular edge–cloud computing,'' \emph{IEEE Internet of Things Journal}, vol.~11, no.~2, pp. 2058--2069, 2024.

\bibitem{intro4}
H.~T. Mouftah, M.~Erol-Kantarci, and S.~Sorour, \emph{Connected and Autonomous Vehicles in Smart Cities}.\hskip 1em plus 0.5em minus 0.4em\relax CRC Taylor and Francis, Boca Raton, Florida, USA, 2020.

\bibitem{IoT-ieee2}
X.~Qiang, Z.~Chang, Y.~Hu, L.~Liu, and T.~Hämäläinen, ``Adaptive and parallel split federated learning in vehicular edge computing,'' \emph{IEEE Internet of Things Journal}, vol.~12, no.~5, pp. 4591--4604, 2025.

\bibitem{intro5}
Y.~W.~Z. Cui and R.~Ke, \emph{Machine Learning for Transportation Research and Applications}.\hskip 1em plus 0.5em minus 0.4em\relax Elsevier, 2023.

\bibitem{IoT-ieee9}
X.~Wang, J.~Lv, A.~Slowik, B.-G. Kim, B.~D. Parameshachari, K.~Li, and G.~Feng, ``Augmented intelligence of things for priority-aware task offloading in vehicular edge computing,'' \emph{IEEE Internet of Things Journal}, vol.~11, no.~22, pp. 36\,002--36\,013, 2024.

\bibitem{IoT-ieee5}
C.~Zhang, W.~Zhang, Q.~Wu, P.~Fan, Q.~Fan, J.~Wang, and K.~B. Letaief, ``Distributed deep reinforcement learning-based gradient quantization for federated learning enabled vehicle edge computing,'' \emph{IEEE Internet of Things Journal}, vol.~12, no.~5, pp. 4899--4913, 2025.

\bibitem{intro5.5}
A.~Matin and H.~Dia, ``Impacts of connected and automated vehicles on road safety and efficiency: A systematic literature review,'' \emph{IEEE Transactions on Intelligent Transportation Systems}, vol.~24, no.~3, pp. 2705--2736, 2023.

\bibitem{IoT-ieee1}
S.~Goudarzi, S.~Ahmad~Soleymani, M.~Hossein~Anisi, A.~Jindal, and P.~Xiao, ``Optimizing uav-assisted vehicular edge computing with age of information: An sac-based solution,'' \emph{IEEE Internet of Things Journal}, vol.~12, no.~5, pp. 4555--4569, 2025.

\bibitem{intro9}
L.~Hou, M.~A. Gregory, and S.~Li, ``A survey of multi-access edge computing and vehicular networking,'' \emph{IEEE Access}, vol.~10, pp. 123\,436--123\,451, 2022.

\bibitem{intro6}
J.~Clancy, D.~Mullins, B.~Deegan, J.~Horgan, E.~Ward, C.~Eising, P.~Denny, E.~Jones, and M.~Glavin, ``Wireless access for v2x communications: Research, challenges and opportunities,'' \emph{IEEE Communications Surveys \& Tutorials}, vol.~26, no.~3, pp. 2082--2119, 2024.

\bibitem{IoT-ieee7}
S.~S. Shinde and D.~Tarchi, ``Collaborative reinforcement learning for multi-service internet of vehicles,'' \emph{IEEE Internet of Things Journal}, vol.~10, no.~3, pp. 2589--2602, 2023.

\bibitem{intro6.5}
Annu and P.~Rajalakshmi, ``Towards 6g v2x sidelink: Survey of resource allocation—mathematical formulations, challenges, and proposed solutions,'' \emph{IEEE Open Journal of Vehicular Technology}, vol.~5, pp. 344--383, 2024.

\bibitem{IoT-ieee10}
Z.~Zhang and F.~Zeng, ``Efficient task allocation for computation offloading in vehicular edge computing,'' \emph{IEEE Internet of Things Journal}, vol.~10, no.~6, pp. 5595--5606, 2023.

\bibitem{parisa2}
P.~F. Moshiri, M.~Simsek, and B.~Kantarci, ``Joint optimization of completion ratio and latency of offloaded tasks with multiple priority levels in 5g edge,'' \emph{IEEE Transactions on Network and Service Management}, 2025.

\bibitem{IoT-ieee4}
W.~Wang, Y.~Zhang, Y.~Zhang, Q.~Liu, T.~Wang, and W.~Jia, ``Online dependent task offloading by application partitioning in edge intelligence for internet of vehicles,'' \emph{IEEE Internet of Things Journal}, vol.~12, no.~5, pp. 4860--4871, 2025.

\bibitem{intro8}
R.~Shen, M.~Gao, W.~Li, and Y.~Li, ``Dynamic task offloading in distributed vec networks: An exploration and exploitation assisted contract-theoretic approach,'' \emph{IEEE Transactions on Vehicular Technology}, vol.~73, no.~4, pp. 5717--5729, 2024.

\bibitem{IoT-ieee3}
Z.~Chen, Z.~Huang, J.~Zhang, H.~Cheng, and J.~Li, ``Resource allocation and collaborative offloading in multi-uav-assisted iov with federated deep reinforcement learning,'' \emph{IEEE Internet of Things Journal}, vol.~12, no.~5, pp. 4629--4640, 2025.

\bibitem{intro10}
W.~Fan, Y.~Su, J.~Liu, S.~Li, W.~Huang, F.~Wu, and Y.~Liu, ``Joint task offloading and resource allocation for vehicular edge computing based on v2i and v2v modes,'' \emph{IEEE Transactions on Intelligent Transportation Systems}, vol.~24, no.~4, pp. 4277--4292, 2023.

\bibitem{parisa1}
P.~F. Moshiri, M.~Simsek, and B.~Kantarci, ``On the interplay between network metrics and performance of mobile edge offloading,'' in \emph{ICC 2024 - IEEE International Conference on Communications}, 2024, pp. 4018--4023.

\bibitem{execution}
H.~Wang, B.~Kim, J.~Xie, and Z.~Han, ``Energy drain of the object detection processing pipeline for mobile devices: Analysis and implications,'' \emph{IEEE Transactions on Green Communications and Networking}, vol.~5, no.~1, pp. 41--60, 2021.

\bibitem{mahsaicc}
\BIBentryALTinterwordspacing
M.~Paknejad, P.~F. Moshiri, M.~Simsek, B.~Kantarci, and H.~T. Mouftah, ``A reliable and efficient 5g vehicular mec: Guaranteed task completion with minimal latency,'' 2025.
\BIBentrySTDinterwordspacing

\bibitem{tasksize2}
J.-H. KO and N.~KIM, ``Performance analysis of object detection models for vehicle-related image services,'' \emph{Journal of Theoretical and Applied Information Technology}, vol. 100, no.~12, 2022.

\bibitem{tasksize1}
H.~Wang, J.~Xie, and M.~M.~A. Muslam, ``Fair: Towards impartial resource allocation for intelligent vehicles with automotive edge computing,'' \emph{IEEE Transactions on Intelligent Vehicles}, vol.~8, no.~2, pp. 1971--1982, 2023.

\bibitem{para1}
W.~K. Seah, C.-H. Lee, Y.-D. Lin, and Y.-C. Lai, ``Combined communication and computing resource scheduling in sliced 5g multi-access edge computing systems,'' \emph{IEEE Transactions on Vehicular Technology}, vol.~71, no.~3, pp. 3144--3154, 2022.

\bibitem{para2}
M.~Gao, R.~Shen, L.~Shi, W.~Qi, J.~Li, and Y.~Li, ``Task partitioning and offloading in dnn-task enabled mobile edge computing networks,'' \emph{IEEE Transactions on Mobile Computing}, vol.~22, no.~4, pp. 2435--2445, 2021.

\bibitem{bandwidthsize}
M.~S. Bute, P.~Fan, G.~Liu, F.~Abbas, and Z.~Ding, ``A collaborative task offloading scheme in vehicular edge computing,'' in \emph{2021 IEEE 93rd Vehicular Technology Conference (VTC2021-Spring)}.\hskip 1em plus 0.5em minus 0.4em\relax IEEE, 2021, pp. 1--5.

\bibitem{guardband}
G.~T. 38.101-1, ``5g; nr; user equipment (ue) radio transmission and reception,'' ETSI, Technical Specification (TS) 38.213, 10 2022, version 17.6.0.

\end{thebibliography}

\end{document}